# ABOUT THE IMPOSSIBILITY OF QUANTIFYING THE KNOWLEDGE AND OF ESTABLISHING CONSEQUENTLY ITS CORRELATION WITH MONEY.

By
José Carlos Bermejo Barrera
University of Santiago de Compostela

**ABSTRACT:** This article shows how the Universities and higher education and research institutions obey more and more to criteria of profitability and effectiveness, following the example of the industry. The Universities and the academic communities are the main producers of knowledge, which is not quantifiable and, therefore, cannot find an equivalent in money, which is quantifiable. Many scientists intend to establish a connection between knowledge and money, using for that a set of criteria of evaluation. Being useless these criteria, the scientists resort to a system of honour, but the science is essentially normal knowledge.

Since the end of the Second World War, when the production and technological innovation capacities decided the last result of the conflict, in the industrialized world an idea has grown of we are living in the called society of knowledge. According to this idea the industry, besides having the production capacity of goods en masse —these goods might be weapons or consumption products—, must be endowed with a great innovation capacity, which allows that a company survives in a market more and more competitive with a massive process of handing over technologies.

The same process affected the Universities, first in USA and then in the rest of the world. They began to think that their teaching and searching functions should be also determined by the criteria of profitability and of development of the technological innovation capacity. These criteria are often formulated in a confusing way, for example the idea of that the University —until recently the main institution producing knowledge— should be to the service of the society. Below we shall try to show how behind these criteria is hidden a wrong epistemological principle, not clearly formulated, which is used as an alibi for another two ideas: the praise of the ideology of market with a neoliberal complexion and the self-justification of the existence of the academic communities. With this aim we shall formulate our line of argument in the form of successive thesis for culminating in the essential analysis of the question to solve.

I

**Thesis 1**: The human knowledge —from now on we'll call it only knowledge— could be defined as a set of statements. These statements make reference to the different aspects of the reality.

But if we add the notion of ontological hierarchy to this epistemological perspective, we'll see that the idea about the no enumerability of the set of statements of the knowledge becomes more feasible.

Nowadays we know that the original idea about the atom like the slightest part of the matter, which comes from Greece, must be corrected. Not only because the atom is composed by a nucleus and electrons, but also because the heavy particles (hadrons) forming the atomic nucleus can be discomposed in quarks (up, down, top bottom), and with that we`ll have a discipline: the Quantum Chromodynamics, which analyses the last components of matter in a level much deeper than Nuclear Physics.

But the quarks could be also analysed, at least theoretically, from the theory of the superstrings. And so, the ontological hierarchy would be:

Superstring

Quarks

Particles

Atoms

Molecules

Avoiding the fact, already quoted, that all these theories depend for their part on another mathematical theories, and all of them on a set of logical and philosophical ideas and principles, we could go on the called by Aristotle the great chain of being in vertical sense.

In the world the chemical compounds exist in the form of molecules and the finite elements of the periodic system are combined in innumerable ways, giving rise to the birth of the emergent properties.

It is well known that the hydrogen has certain properties and the oxygen others, but when the hydrogen and the oxygen combine in a determined proportion forming the water, the properties of the last one are not included in those of the hydrogen or of the oxygen.

A principle of the quantum mechanics and the theory of relativity is: there is not a spatially privileged observer and on the quantum level the properties of the instrument of watching modify the properties of the particle observed, whose existence is often induced or provoked, which cannot be made if the particle is confined into other one.

In order to understand ontologically how is a physicist into a laboratory putting aside the properties of the studied object, we must establish the next ontological hierarchy:

A physicist is a body composed of chemical elements forming an unknown and innumerable number of compounds. Those compounds can be organic and inorganic. The organic compounds form the skeleton of the units of life: the cells. The chemical compounds of life are governed by genes, which form innumerable combinations from the four bases of their composition: adenine, cytosine, guanine and thymine. Each cell has genes, but the cells come together forming tissues, the tissues organs, and the organs are linked up anatomically through physiological mechanisms of chemical character.

In the physicist's body there is also an anatomical and functional hierarchy, and on the top of this would be the brain made up of 100,000 millions of neurons linked each other by thousands of synapses and with a working regulated through neurotransmitters. This makes possible that the physicist formulates his statements. But we must take into account that the set of emergent properties grows exponentially from the atom to reaching the stating of the thoughts.

In this analysis we are not taking into account the equipment for the physicist makes his searching, tools often infinitely complex, like in the case of a particles accelerator. From this example we think that it could be clearly guessed that the knowledge is composed by a set of statements linked up through millions of channels, in the same way that the neurons are linked by their synapses. Because of that it is very difficult to know how many assumptions are hidden in a certain moment behind the formulation of a statement which apparently makes sense isolated, while it refers to some aspect of the reality.

Against this argument could be put forward other one: There are many scientific disciplines academically consecrated that seemingly enjoy a certain grade of autonomy.

Each one of them works with a set of statements much more limited, which seems to be numerable. In spite of it, we can even support the thesis 1, widening it with the thesis 2.

**Thesis 2:** There isn't a scientific discipline constituted in such a way that the set of its statements has the properties of absolute completeness and connectivity, which is a consequence of the Gödel's theorem.

Gödel formulated his theorem applied to the formal systems, proving that in an axiomatic system can always rise unspeakable propositions into the system. If that occurs in an axiomatic system, where the grade of coherence cannot be changed by the limitations or the mistakes of the searching —unavoidable in all the empirical sciences—, it will happen much often with all the sciences referred to the reality; or in other words, those not merely formal.

Although the scientists think that their knowledge is autonomous, that is not always true. It is not a property of that knowledge, but a subjective perception of the scientist, determined by his academic or professional position or by his place in the market.

All the scientists are qualified, more or less, for formulating statements about the sector of the reality studied by them, but not for formulating statements about their own knowledge, because in this case they would be statements about other statements, and the method for formulating them is not that of a science related to a sector of the reality: atoms, molecules, cells...

In fact, some scientists, like Edward O. Wilson (E.O. Wilson, 1998), speak again about the unit of knowledge, claiming a main principle of the positivist and neopositivist philosophy, but without being aware of it.

Let`s see an example. A neurologist may consider that his discipline is autonomous, because it studies a certain cell: the neuron. Nevertheless, it is well known that the field of neurosciences is highly complex. It covers not only the anatomical level, but also the genetic and the biochemical ones, due to the importance of the neurotransmitters and the relation of these to the endocrine system. Not to mention the cognitive, psychological and emotional aspects.

However, all this could be avoided in order to the next reasons: academic blindness and business interests. For instance: Ronald Davies, director of the Baylor College of Medicine of Houston (USA), stated in the journal *El País* (29-06-2005) that he was studying the genetic and molecular basis of brain diseases, like the Alzheimer and the schizophrenia.

His method involves researching with the «vinegar fly» through electric shocks. If those shocks modify the behaviour of the fly, the changes in the genes and the neurotransmitters must be analysed for synthesizing a molecule marketable as medicine.

A problem rises here, pointed by the journalist interviewer: the reductionism of the method used.

It is clear that the brains of a vinegar fly and of a human being are very different. Not only for the infinitely lower number of neurons of the fly, but also because its brain is not structured like our brain, in which is essential the development of the cerebral cortex, indispensable in the language and in our cognitive powers.

Not only that. Moreover, the schizophrenia is a pathology so complex that even some psychiatrists speak about its future vanishing as clinical entity in the DSM, because the symptoms of the catatonic schizophrenia and the paranoid schizophrenia share seemingly few characters in common.

The paranoid schizophrenics listen to voices, and not the vinegar fly. The fly cannot have visual illusions, sex or affective disorders trigged by those electric shocks applied by our investigator.

It is known that no theory explains satisfactorily the schizophrenia (J. Garrabé, 1992), but everyone accept the importance of the social and family factors, also those of emotional kind, and even the historical factors, because this disorder is not registered in the primitive communities.

Davis says that he does without these factors because they are very difficult to analyse, while the biochemical one is easy. In this case he breaks the main rules of the scientific research. ¿Why is he doing it such a thing? He does it, among other reasons, because «discovering» and patenting a molecule for «healing» the schizophrenia or the Alzheimer —in this the emotional or social factors don`t play any role— is a business of enormous magnitude for the multinational pharmaceutical corporation that patents it. Obviously that molecule won't heal the schizophrenia, but if it makes that some symptoms disappear, like gets the haloperidol, its use would be already justified, although its secondary effects trigger Parkinson. The sufferings of the schizophrenics are partially relieved, they recover sociability and, most of all, a big business is made, because there are millions of schizophrenics in the developed world, which can afford to pay for those psychoactive drugs.

Some people will say that the example is perhaps exaggerated. It could be, but it is not invented, but something real and released in a newspaper of big circulation. With the example we have brought up another factor, the economic factor, in other words the money. But that factor is considered inseparable from the production of knowledge in our world, not only by the businessmen, which play their role, but also by the politicians —with a duties over the economic sphere—, and by the scientists —in the same case that the politicians—.

Before going to the formulation of our next thesis, referred to the money and the scientific communities, we must add a complementary thesis, widely known and formulated by T.S. Kuhn in a classic work into the philosophy of science: *The Structure of Scientific Revolutions*.

**Thesis 3** (or T.S. Kuhn's thesis): There are two kinds of progress of knowledge: a) the cumulative growing into a given paradigm, and b) the progress of knowledge due to a total reorganization of the actual knowledge, or just the same, due to the shaping of a new structure of the knowledge.

From this thesis 3 we must develop the thesis 4.

**Thesis 4:** The growing of the knowledge into an established paradigm is only measurable, but not necessarily quantifiable. On the contrary, when there is a change of paradigm we can say only that the set of the statements of the paradigm 1 must be a subset of the statements of the paradigm 2. Sets equally not countable, although one could be the subset of the other one.

An so, a chemist can say that the nowadays periodic system has more elements than a hundred years ago, in the same way that there are more stars in the flag of the USA, we know more proteins now than before or we have synthesized thousands and thousands of organic compounds. Although all this is certain, our previous thesis will be also valid, because chemical theory depends on the Physics and this on the Mathematics, and so on. According to this, we could formulate the thesis basing on a statement that apparently contradicts us.

**Thesis 5:** The growing of the knowledge, into a given paradigm or into several of them, don`t allow that the set of knowledge becomes countable, because the growing in the number of statements implies the growing of the connections among statements following a geometric proportion.

The statement of this thesis is easily understandable knowing only the principles of simple Arithmetic. Something similar is argued by E.O. Wilson in the quoted book, although he wouldn`t share the thesis holding that the set of the statements of knowledge is not countable.

Let`s see an example. The synthesis of the urea allowed break the barriers between organic and inorganic Chemistry, the discovery of the ADN allowed to bring the field of the life closer, excluded for a long time from the scientific determinism, from the sphere of Chemistry. Even the discovery of the autoreplicant proteins, the prions, has brought these fields closer. This unification of fields is not merely reductionist. Or, in other words: from the unification of two fields is not derived the decrease in the number of statements. On the contrary, that number grows, and when growing, also grows the number of interconnections, which gets us away from the countable ideal.

It has been, not proved —because the proofs are only feasible in Logics and Mathematics, but shown that the set of the statements making up the knowledge is not countable.

Let`s go to the second part of our line of argument, in which comes into play the second of our main features: money.

II

We`ll begin putting forward a thesis admitted by any person with a minimal education.

**Thesis 6:** The set of the money is a countable set. The present amount of money must be finite.

It is obvious that money fulfils three functions:

a) It is a means of interchange.

b) It is a pattern of measurement.

c) It is a means for cumulating richness.

At first some formal similarities could be formulated between money and knowledge. Gottlob Frege said that the scientific knowledge is like a passbook: the more money you put into it, more money you have and more profitability you get. Knowledge and money would be in that way cumulative. The problem is that the human knowledge, although can be assessed, as more or lesser, is not quantifiable, as we have stated above. And the capacity of communication of scientific knowledge depends on the capacity of the human communication through the language, which is a social fact.

Actually, when we talk about the relation between knowledge and money, we are talking about a very specific subject, the subject of the output of the capital in which would be combined money and knowledge.

In the social process of production of a good are needed: a) raw materials, b) means of production and c) labour. Each element implies a certain kind of knowledge. Namely:

It is necessary developing knowledge for knowing, spotting and using the raw materials. Secondly, knowledge is required for developing a technology for the industrial processing of those raw materials. And finally, the labour must have a degree of qualification, increasingly bigger, in order to make the productive work, which implies a process of training.

Besides that, since the prices are formed basically into the market from multiple factors, the businessman must have information for developing his production strategy tending to: a) finding markets, b) discovering cheaper raw materials, c) developing a technology for reducing production costs and d) persuading the workers to change their work for the wage paid for him. In this last point it is necessary a strategy of persuasion and consensus. That strategy is not only a thing of the businessman, but depends on the whole social, legal and political system.

A society is not only a means of production, or in other words, a market. On the contrary, a historically determined society consisted of:

a) A means of production in order to produce and exchange the set of goods indispensable for the physical and social survival of the group.

b) A means of constraint in order to repress and punish the external and internal enemies of the system. That way of constraint, made up of army, policy and legal system, civil and criminal, doesn`t look for the profitability, but for the effectiveness in the social and political control.

It may arise two societies with two different means of production —Germany and the USSR between 1933 and 1945—, but with two identical means of constraint: an only party, political policy, abolition of the individual rights, system of concentration camps. Or in a more recent example: the ultraliberal economy applied in a certain moment by the military dictatorships of Latin America could be similar to the economy of the USA, but in the USA there was not a military dictatorship.

c) A means of persuasion, called ideology by Marx. The teachers take part in it, and in its development there are essential disciplines, like the Law, the Political Science, the Sociology, the Economy and the called Human Sciences, as says W.G. Runciman (1983; 1989), designer of this tripartite system.

Not recognising the coexistence of these three systems in order to arouse the production and the market is based on a mistake, pointed by José Manuel Naredo (J.M. Naredo, 2003), which we would call paralogism of the absolute quantification and formulate in the next thesis, recollecting the thoughts of this author.

**Thesis 7:** It is impossible to make a comprehensive economical assessment. The economical assessment has only a relative value, either in the sphere of a specific means of production or in the generic estimates of macro economical magnitudes, which cannot explain the most of the actual economical processes, but merely determine them.

Undoubtedly a businessman assesses the profitability of his productive process. But in that assessment is rejected a set of magnitudes referred to all the components of the productive process. Namely:

a) The businessman assesses the cost of the raw materials forming direct part of the productive process, but many natural resources attached to them are rejected. In fact, only recently has been taken into account the economical impact of the industrialization. This impact may be compensated with the payment of environmental taxes by the tax system. But even so the cost of all the raw materials would be never covered by the producer businessman: water, air could be polluted. In fact, we know that a big part of the atmospheric pollution comes from the economy of USA, a nation not signatory of the Kyoto protocol.

b) Besides, the businessman doesn`t pay the investments in the knowledge necessary for the development of the required technology.

The technique has a history cumulative indeed. The engines and the machines improve gradually their performance. And there is not technique without an applied science, or an applied science without sheer science. And so, we can say that, as the businessman using many natural resources not paid by him, in the same way he couldn`t ever cover the costs of production of the knowledge implicit in the development of the technology inherent in the productive process. In both cases the businessman is a privileged citizen. Not only for this, but also because he uses public resources, roads, trains, ports, in his productive process without paying the corresponding costs.

c) The businessman makes sure his capital grows every day for investing it cyclically. However, the growing of the labour is free. The labour is provided by women, as points Claude Meillasoux (1975), the families pay the costs of the breeding of children and the State supplies their education. All of this is costs of the productive process not assumed by the businessman.

And finally, in the development of his strategy of looking for markets there is a big part that depends on his skills. But also in this case are useful the geographical, geological and scientific knowledge developed by the society in which he lives.

And so, the economical assessment is only a useful fiction. It is a necessary tool to develop a complex productive process, and it explains only partial aspects.

The function of knowledge in the productive process is equally a very limited function. It is a mere tool or another part in that process. In the productive process the knowledge is subordinated to the finality of the process, namely, the production of a profitable good. The knowledge used in a productive process depends on the set of the knowledge. In the same way that the production depends on the existence of a social group, the use of the scientific and technological knowledge in a certain productive process depends on an overall knowledge, result from a social and historical process in which that knowledge has been formed.

Because of that we can say that the businessmen couldn`t pay the costs of their productive process in the sphere of the raw materials, the technological development, the reproduction of labour and the production of the implicit knowledge.

The businessman, and the scientist or the technologist working for him, are privileged citizens, because they benefit from a series of goods that the set of the society produces in a much more degree than the rest of their countrymen.

In the use of the knowledge in the process of production take part the next values:

Profitability

Effectiveness

Power

And all of them are interrelated by a magnitude, money, according to the next scheme:

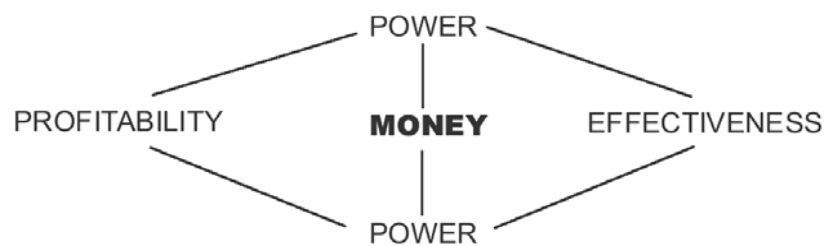

The profitability is merely economical. The effectiveness is material, in the production of goods or in the development of machines of destruction and repression, like weapons. Both of them are indispensable aspects of power and obey to money, because in all the cases finite resources are being used to get certain aims.

We cannot deny the existence of these values, but it is equally true that there are also others, like the political or moral values or the value of knowledge per se. If they are sometimes mixed up is because there is a specific political strategy in the present moment. Many scientist and technologist take part in this strategy, either because they share also those values or because they think that those values reinforce their academic values.

After the crack of the economical and political system of the USSR and its satellite nations, the ideology of the unique thought became established in the Western world; and according to this ideology there is only a feasible economical system: the capitalism, and a political system: the parliamentary democracy; and all of this is supported on the ideology of the human rights.

The decline of the USSR meant the crack of a system of production, the directed economy, of a system of military and police constraint and of a system of persuasion. It became clear that the capitalist economical system is more profitable and efficient than the called socialist. Secondly, it was revealed that such a system was based on a system of constraint not wanted by the majority of population, and so we could say that the way of persuasion also failed.

The fail was so enormous than the ancient USSR didn`t even survive as a way of constraint and its military power dropped due to the economical failure, and the USA stood as the only hegemonic military power in the world.

Only the Islamic Fundamentalism arises nowadays against the American military hegemony and the capitalist and liberal democratic system. The Islamic Fundamentalism doesn`t entail a new means of production, it cannot develop a system of persuasion through rational lines of argument in order to convince the non-believers and it cannot introduce a new system of constraint, but only destabilize the prevalent means, reaching some extremes truly dangerous.

Superpowers like China, that keeps the previous way of persuasion and constraint, accept the capitalist way of production and they are not able of giving, through its way of persuasion, lines of argument for refuting the unique thought.

Nevertheless, from the fact that the Capitalism is more efficient and profitable than the Socialism is not inferred that the Capitalism is necessary. Going form the «is» to the «must be» is known with the name of *naturalist fallacy* in the ethical theory. We cannot think that because in Spain a husband kills his wife almost each week, this must be made because it happens. The same occurs with the Capitalism and with the theory proposing the subordination of knowledge to the process of economical production.

In fact, the world economical system —whose complexity in financial, trading and productive webs is so big than it cannot be processed informatively— arises serious problems, which points its unfeasibility. Let`s see some examples:

If the level of consumption of energy and of the mechanization of the USA becomes more widespread through the Planet, which should happen, if its economical system is universalizable, not only the fuels —fossil or of other kind— would be insufficient, but also the atmospheric pollution would make impossible the life on the Earth.

The present world, simplifying the subject with explanatory purposes, can be divided into a highly industrialized world living in the sphere of the «society of knowledge» and a not industrialized word, or at least much lesser industrialized.

We could guess that world supplying its agricultural production to the industrialized world. But it happens that the industrialized world invests six times more money to protect its own agriculture than to promote the development of the called Third World. The European Union invests, for instance, the half of its Community funds in subsidizing its own agriculture, in which works only the five per cent of its population.

The industrialized world transfers its industries to the world 2, called so for convenience. But it happens that the wages of the workers of the world 2 are so low than their purchasing power cannot contribute completely to the development of that industry. And similarly, part of the people of the world 1 becomes poor and his purchasing power drops. This points to the existence of a serious structural problem for the survival of that economical model.

In the world of the «society of knowledge», in which the 70 per cent of the worldly population has never phoned, we hardly could connect them to Internet. There are countries like Afghanistan where only the five per cent of people has running water. The consumption of water for farming and domestic uses is a worldly economical problem of first rank. Without enough water and without hygienic systems it is impossible keep an acceptable health situation.

In the world of the «society of knowledge», the physicians and the pharmaceutical industry research hardly the catalogue of more or less five thousand rare diseases, because it is not commercially profitable. The poor people`s diseases, like the Chagas disease, has not any interest. The society looks for medicines for rich people, like the statines or medicines for Alzheimer, highly profitable because of their high prices.

The inner combustion engines waste almost all the energy that burn. We don`t make others, because we don`t know how —which would show the limited character of our system of knowledge—, or because we don`t want and it is interesting for the oil industry, which is worst.

We may state that our industry scientific system has many problems. It is not spreadable all around the world. That system is increasingly more powerful. Nowadays we can say, like points John Ziman (J. Ziman, 2000), that the science lives in the postacademic stage. In Japan, for instance, the 70.5 per cent of searching comes from the industry. Some times ago Pierre Thuillier emphasized the military, political and economical manipulation of science through many examples.

Obviously the industry needs research, but the logic thing is that pays for it, as pays for the building of its plants. All the knowledge cannot be subordinated to the production because of two reasons. First, because it is impossible. The knowledge is a very complex system, and nobody can establish authentically the border between the pure knowledge and the applied one. The industry only funds the applied knowledge, but it gets benefits also from the pure one, funded by the State, which is not very fair.

Subordinating the knowledge only to the industry is unfair and also counterproductive. The engineers use to be mentally conservative and tend to repeat the acquired knowledge. We could say that a guarantee for the progress of knowledge is that an important part of its study is not controlled financially by the industry.

The ideology of market appeals to liberalism and to the defence of the rights of individuals. But it appeals to a kind of liberalism called *possessive liberalism*, which privileges the right of property over many others.

Of course, the right of property is one of the human rights. But we must introduce a nuance. Every right calls at least for the existence of two persons, A and B, and so the right of A only exists when recognized by B, and vice versa.

However, in the case of the right of property we must introduce a nuance pointed already by Marx: the private property of the means of production implies not only the recognition of the bussinesman`s right from the workers, but also the fact that in many cases the worker`s survival depends on the wages paid to him by the businessman. And it is usual that many workers accept a job, not because the businessman persuades them, but because they have got no choice.

The liberal thinking in the nineteenth century considered that the unions must not exist —because they limited the freedom of recruitment— neither the social advantages: health, unemployment and retirement, because they slowed down the growing of the market and contributed to the survival of the less apt, as pointed Herbert Spencer, a philosopher and sociologist that prophesied that the development of the industry would put an end to the war, evolving from the military society to the industrial one. The History of the twentieth century and the two World Wars refuted his assertion.

If nowadays we can think in a different way than in the nineteenth century in these subjects is because another values —political, social and moral values— slowed down the development of the pure market economy. An economy with a purity qualified by the economists themselves.

The task of the natural or social scientists and of those working in human sciences is contribute to the progress of knowledge, considered like a common good, and from it contribute to the formation of a rational public opinion in a democratic context by means of the spreading of that knowledge.

Jürgen Habermas pointed some time ago that it cannot exist a democratic society if there is not a process, more or less free, of formation of public opinion (J. Habermas, 1962). One of the basic tasks of the scientists and intellectuals is contribute to the formation of that process, as says Jeffrey C. Goldfarb (J.C. Goldfarb, 1998), which is not easy in a world with the most part of communication companies private and where the political parties intend to monopolize the formation of public opinion.

The social production of knowledge can be determined by the criteria of profitability and effectiveness, but it must obey also to other political and moral values and to the search of the knowledge, a value itself. The companies look basically for the profitability and the effectiveness, and the State, which supports public health systems, infrastructures and the national defence system, must take charge of promoting the pure knowledge and the political and morally necessary knowledge not profitable, but essential in the development of the way of persuasion.

The way of persuasion must take care of the production of a kind of knowledge, which task is creating the social consensus. That knowledge can be ruled by democratic systems of values recognizing the human rights, or not, like happened in the Fascism, The Nazism and the Stalinism. In the case of nations with democratic constitutions, the social promotion of that not productive knowledge obeys to the constitution and must be mostly a task of the State. And the same happens with the spreading of those values through the education.

We can synthesize our proposition, from the thesis 1, in the next way:

**Thesis 8:** The human knowledge can be defined as a not numerable set of statements.

That set includes several subsets, intersected among them.

Some of those subsets can be more linked to a limited number of values than to others (for example: the technology of the inner combustion engines is interrelated to effectiveness and profitability). Nevertheless, on a global level:

The set of statements that form the knowledge is linked to a complex system of values where the values quoted above, profitability and effectiveness, must be subordinated to other values with bigger regulatory hierarchy (the political and moral ones). It is impossible establish basically a correlation between knowledge and money, because the set of knowledge is not numerable and the set of money yes.

We had said above that the State must promote the knowledge and it must exist institutions devoted to its production and teaching; among them, it plays an important role the University.

Let`s see now if the University can fulfil that function politically and if the own structure of University makes easier or difficult this task.

III

The European Universities were born in the Middle Ages as corporations for the education of jurists, theologians and, to a lesser extent, physicians. They followed in that way until the first part of the nineteenth century, when it began a paradox —because the development of the modern science had taken place basically out of these institutions—, that some of their more distinguished professors pointed an essential contradiction in the own university institution. As said I. Kant in *Der Streit der Fakultaten* (I. Kant, 1983), it was curious that the Faculty with a higher hierarchy in the knowledge, Philosophy —in times of Kant it included all the sciences—, had the lesser social prestige.

The University was also a centre for the education of the elites. One went to Oxford or Cambridge in England in order to be trained as a clergyman or as a lawyer, or to obtain a label of social honour, which was pointed by Cardinal Newman (C. Newman, 1912) even for the second half of the nineteenth century. The access to those two universities was exclusive for the sons of the nobility or the upper middle class who, more than learning, spent there a time.

The kantian idea of favouring the production of knowledge as mission of the University was defended in the twentieth century in Spain by José Ortega y Gasset (J. Ortega y Gasset, 1940), who foresaw the danger of turning the University into a centre of education of technicians, falling so in what he called the «barbarism of the specialism».

But not only the unstoppable progress of the science-technique foresawn by Ortega sixty years ago can mean a limitation in the mission of the University understood as centre tending towards the production of knowledge, but also can contribute to it the structure of the academic communities.

The university professors had been the target of interesting sociological analysis by authors as Pierre Bourdieu (P. Bourdieu, 1984); Tony Becher and Paul R. Trowler (T. Becher and P.R. Trowler, 2001) and Bill Readings (B. Readings, 1999).

From their analysis we could formulate the next thesis:

**Thesis 9:** the University and its scientific communities are the most suitable institutions to produce knowledge freely, but there are psychosocial hindrances making difficult this task.

The existence of these hindrances, linked to the economical and political pressure, could make that task impossible. And so it happened that:

The University would be an institution created for developing a task obstructed by the creators themselves. The existence of the University would be a contradiction in the terms.

Pierre Bourdieu said (P. Bourdieu, 1984) that the *homo academicus* lives in a double system of desire, swaying between the *libido sciendi* and the *libido dominandi*, the two forces governing the human soul, according to Saint Augustine.

The university men want to know. Knowing is pleasurable for them and they devote their lives to its search, learning skills and abilities to create it in the frame of their speciality and developing techniques of exposition of them, according to the rhetorical keys of each speciality, analysed by Alan G. Gross (A.G. Gross, 1990) in the case of the rhetoric of science.

But the *homo academicus* feels also a passion for power. Plato, a pure philosopher, wanted to govern and advised a tyrant, who sold him as a slave. In his political utopias the city must be governed by the philosopher-kings.

With the Christianity the philosophers` guidance function was assumed by the priests, who are the producers of the essential knowledge: the Theology and they advise the governors threatening them with the throwing out the Church.

The catholic and protestant Inquisition, as in the case of Calvin, intended to slow down the development of the free thought and of the science, without getting it. The science will triumph, the *philosophes* of the Enlightenment will become its champions, but also they, as Voltaire and Diderot, will want to be —and they will be— advisors of the kings and queens. The Positivism of Comte preaches in favour of the government of the scientists and the businessmen. And this vindication is assumed partly by the socialist thought, in which the governing mission in Politics must fall to who knows the «scientific laws» of the development of the society and history.

That passion for power is shown in two aspects: a) the academic power and b) the economic and political power, until now secondary.

The university people, according to Bourdieu, cumulate a *symbolic capital*. That capital is formed by an accumulation of honours, from the publications: books, articles, discoverings or patents, until the accumulation of «external signs of knowing»: membership of scientific societies, chairmanship of them, attendance to meetings and congresses...

Becher and Trowler divide the professors into monks and courtiers; the monks are more oriented to the *libido sciendi* and the courtiers to the *libido dominandi*, although between the two categories there is not an absolute division. In the same way, Readings divide them into countrymen and urban men, sharing the same characteristics.

The existence of the academic institution itself and of the scientific communities makes easy the production of the scientific knowledge. However, it also stops it, because both institutions claim for the adaptation to some patterns of behaviour and forms of expression and thought that can be, or not, dictated by the nature of the object studied.

T.S. Kuhn pointed the epistemologically conservative character of the scientific communities. If to that conservative character intrinsic to them, we add the desire of subordination of them to some researching parameters dictated by the profitability and the effectiveness, then its grade of freedom of thought will be minimal.

The university man of Bourdieu looked for «symbolic capital», honours confirming him as a wiser. He was prepared to drop a part of his intellectual and personal freedom, because the satisfaction attained was bigger than the sacrifice assumed.

The new *homo academicus*, twenty years after the publication of the Bourdieu`s book, submits in order to get the symbolic capital and also submits to the business and political interests in order to get «money capital». And so, the university man not only submits to his scientific community, but to the interests of the companies and of the public institutions responsible for funding the research.

The researcher must not only find the scientific recognition of his fellow men, but also the confidence of the state and business assessors for funding his research.

Those assessors, controllers of the development of knowledge according to external criteria —economic or political—, could be also scientists champions of those limited values, or old scientists who don`t practise yet the research because they have come to the conclusion that the only *libido* giving pleasure is the *libido dominandi*.

The state or company «assessors» may become the new theologians, adoring the profitability and the effectiveness, or the new inquisitors. Their mission is to develop mechanisms of evaluation more and more complex, which justify their existence as corporation, in the same way that the Egyptian scribes complicated purposely the art of handwriting to justify their privileges, derived from a long learning. Or also the Chinese mandarins considered as a highest art the calligraphy, complement of the knowledge of thousands of rules and orders.

These mechanisms of control of the research are partly necessary, because the resources are limited. But it is also true that they are conditioned by the managerial and neoliberal ideology, forgetting that we cannot give exclusiveness morally to a kind of values, because all the values form a connected system, where it is very difficult to foresee the consequences of the violation of a certain value, as points Keith Graham (K. Graham, 2002).

Sometimes the managerial ideology intends to decide the existence of institutions that cannot be companies due to its own character, like in the case of the University, the health institutions or the armies. Almost all the health investment obeys to ethical reasons. ¿What is the profitability of preserving the life of old people, which is very expensive, when the benefits of pharmaceutical industry increase the public deficit? ¿Why are there nuclear weapons? Basically for not using them, because the nuclear strategy is based upon the deterrence of the enemy, whom «I convince» that a mutual attack would cause our mutual destruction.

The managerial conception of the University is a contradiction in terms, because it means denying the existence of the own University and its basic purposes. That conception hinders or prevents the development of the knowledge. And if we are consistent with its implementation, we should ask the closure of the University, as says Readings.

We must add to all this that who defends this conception into the academic institution is fooling himself and deceiving the others, because there is not real businessmen in University. They are using public resources and government property, don`t pay taxes and, moreover, employ highly qualified labour in working conditions of underemployment.

The establishment of the unique thought and the neoliberal ideology has a clear conservative character and, as it is logic in this kind of thought, conveys a certain panglossian attitude, according to which we live in the best of the possible worlds. The assumption of this perspective by the scientists has very definite consequences. First, they are not aware of the limitations of the scientific thought and consider it as a closed and perfect system. Secondly, just as Leibnitz wanted —Voltaire caricatures him through Pangloss—, all the possible things are already effective or real.

Given that these scientists identify themselves with the current political and economic system, they will formulate unwittingly the last of our thesis, although of course in an affirmative sense, not negative, which is the real one.

**Thesis 10:** The political and administrative process of assessment of the knowledge can recount the value and the development of the scientific knowledge.

That is not true. Let`s see why.

The scientists imbued with the managerial ideology —they are convinced that it is possible to establish a clear connection between knowledge and money— imitate the managerial reckoning in their processes of assessment.

We have seen above that the economic assessment is very limited and has only validity in the design of a process of production, when it is made the estimate of costs and assessed the possible profitability. That assessment is not perfect, leaving aside the characteristic limitations of the economic reckoning, because the mechanisms of formation of prices are conditioned by many factors, some of which are not merely rational, but sociological, psychological and political. All of us know the sensitiveness of markets to the political tenseness and the possible warlike threatening. But, besides that, the success of a product in the market —moreover, if it is not a staple item with a non-elastic demand— is conditioned by the likes of the consumers and by the fashions.

The businessman can design a process of production, but the complex and extensive reality of market makes that, through the play of many interacted factors, his assessment leads to the success or to the failure. We could establish a simile. If we say that the businessman is an internal observer, because he starts from a subjective perspective, although he is endorsed by many data and techniques of interpretation, and we call external observer to other one, who analyses the process, not before but after the manufacture of the product and its destiny in the market, we must admit that the righter perspective would be that of the second observer. In the same way that in a historic process who analyses a process after its ending has a righter perspective than who is living it, although this one has a direct experience.

Given that the process of assessment of the scientific knowledge imitates the managerial logics, we must admit that this process share the same limitations, but the scientists assessors will not recognize it, because they have a panglossian and all-embracing view of the science.

Those assessors can only assess the value of a process of research if this is oriented to the manufacture of a product for the market. And that occurs because the profitability is an economical concept. Or they can consider the more or less effectiveness of a machine.

What they cannot make is quantifying something not numerable.

The most of the scientists use to be very ignorant in the history of their discipline, whose study is set aside for the professors of «History and Philosophy of Science». Because of that they are not aware of two essential facts: 1) A science is the result of a historic process, where the acquired knowledge is considered an observed fact and the accumulation of knowledge allows, precisely, the advance of the knowledge because it must not be proving always the basic truths on which stands each scientific domain. 2) Like before it was said «*habent sua fata libelli*», the success of a scientific theory cannot be predicted before its developments. The perspective of future will allow check that success or its failure. The scientists can predict events if they have some mathematic law —as in the case of the eclipses—. What they cannot predict is the future of their own discipline. As we have seen above, a scientist can speak about the section of the reality he must study, but not about his science, and even lesser about the science in general.

If we put aside the process of assessment linked to the production of a good or a more or less effective instrument and we focus on what we could call the mere knowledge, we will see that the criteria of assessment used are purely external. In this second case they are not corrupted by concepts of the economic theory, but by the institutionalized system of academic honours, which forms what we have called — following to Pierre Bourdieu— the «symbolic capital».

Let`s see some examples of it, of curious character.

It is generally admitted that the research can be measured through the number of projects of research of an institution or of a researcher. But this is a nonsense, because we should quantify first, not the number of projects or its cost —another curious magnitude admitted as a symbol of quality—, but the quantity of knowledge produced. That quantity cannot be measured, because the knowledge is a not numerable set of statements.

The knowledge, besides developing in relation to the spheres of the production and the technology, is also developed in the sphere of values. There is a hierarchy of values, but not a quantification, and despite of attempts like that of Javier Echevarría (J. Echevarría, 2002), it is not possible establish a method of scientific assessment where it was attained a quantification, which can be only monetary starting from the values.

The discussions about the hierarchy of values are political or ethical discussions, but no means they are economic-managerial.

If we put aside this curious quantifying of knowledge through the number of projects and its monetary value, we`ll see that the other criteria of assessment used are symbolic or honourable.

It is obvious that the scientific knowledge is produced massively. There are thousands of scientists in the world and dozens of thousands of articles are published, to the point that we need computer tools in order to find among them those interesting for us. However, the scientist wishes recognition, needs stand out in that mass of anonymous scientists.

The science is basically normal knowledge (T.S. Kuhn, 1962). The scientific revolutions are very few and the number of exceptional scientists —personified symbolically by the public opinion in the figure of Albert Einstein— is minimal. And so, it is necessary to develop institutional mechanisms of hierarchic promotion, like the arranging of magazines according to categories and the resulting acquisition of more honour when publishing in the more important magazine. The posts in the scientific societies, formed by hundreds or thousands of members, are another sign of intellectual recognition. Or the ritual representation of acts granting honour in the appropriate institutions, like the attendance at congresses or the hierarchical involvement in them: with papers, communications or chairmanships.

The odd thing is that all these honours are also quantified. And so, the symbolic capital of the scientist would be, for example, the sum of his publications, scientific-honourable posts and involvement in scientific acts, like congresses and conferences. But these magnitudes cannot be summed. And the contribution of the scientist to the increase of the knowledge is not measured multiplying a constant by the number of his publications. Such a thing would be foolish and everyone could see it, and it would be also impossible, because the knowledge is not quantifiable.

The big advances in the knowledge don`t happen only through the quantitative accumulation of information, although this is essential, but due to the introduction of new systems that reorganize the whole of a field of knowledge. Those structural reorganizations are known as «scientific theories» or «scientific revolutions» in the kuhnian terminology. It is generally admitted that their number is very short, their emergence cannot be predicted and nobody can plan or develop investigation programs for produce them, because their emergence is conditioned by the growing itself of the scientific knowledge, but also by external factors: philosophical, social and even political and psychological.

A project of investigation only can be established into the sphere of an established theory. Consequently, the advance of knowledge only can be planned in the spheres that lesser academic glory bestows on the scientist; in the fields where the keynote is to be normal, to comply with the established patterns and not to stand out. But the scientist has the psychological need, promoted by the own competitiveness of the academic world, of standing out. In order to that he will create this system of honours where external signs are mixed up with the production and the assessment of the knowledge. And so, the scientist finds himself immersed into a community that plans his intellectual life, sometimes also the affective one, and he loses the critical sense against his own discipline and against the academic system to which he belongs. He can even lose, as it happens in the developed world, the political perspective when he worships, in addition to the idols of the tribe, as said Sir Francis Bacon, the idols of the forum, in this case the *Forum Boarium*, the square where the business deals were made in the ancient city of Rome.

In the seventieth-century lived in Holland a Jewish philosopher, Baruch Spinoza, who was alienated as a Jewish and alienated twice when he was thrown out from the Synagogue because he had dared to think freely about the Jewish religion and its sacred books. Spinoza had a big passion for knowledge, represented for him by the Mathematics. He said that the only possible love for God was the love for the knowledge. To prove it, he wrote a treatise, *Ethics*, in the form of mathematical postulates. We wish remember now the last of his postulates, in which the whole treatise culminates; nowadays this postulate is of paramount importance, because the scientists are not able of seeing it. Let him speak:

«But all things excellent are as difficult as they are rare».

Baruch Spinoza, *Ethics*, XLII. Scholium.